\documentclass[11pt]{article}
\usepackage{blois,epsfig}

\def\be{\begin{equation}}
\def\ee{\end{equation}}
\def\bea{\begin{eqnarray}}
\def\eea{\end{eqnarray}}

\begin{document}
\vspace*{4cm}
\title{Asymmetric WIMP Dark Matter in the presence of DM/anti-DM Oscillations}

{
\author{Gabrijela Zaharijas}
\address{The Abdus Salam International Centre for Theoretical Physics,\\ 
Strada Costiera 11, I-34014 Trieste, Italy}
\vspace{0.5cm}

\maketitle\abstracts{
The class of `Asymmetric Dark Matter' scenarios relies on the existence of a primordial particle/anti-particle asymmetry in the dark sector {\it related} to the baryon asymmetry as a way to address the observed similarity between the baryonic and dark matter energy densities today. Focusing on this framework 
we calculate the evolution of  the dark matter relic abundance in the presence of particle/anti-particle oscillations. We show how oscillations re-open the parameter space of asymmetric dark matter models, in particular in the direction of allowing large (WIMP-scale) DM masses. 
Finally, we constrain the parameter space in this framework by applying up-to-date bounds from indirect detection signals on annihilating DM.}

\section{Introduction}

In the last years we witnessed a revival of interest in Asymmetric Dark Matter (aDM) models (for a recent list of references see \cite{Cirelli:2011ac,Tulin:2012re}) in which dark matter (DM) abundance is determined  {\em by an initial asymmetry} between dark matter particles and anti-particles. The main motivation for this approach is to address the observed similarity in the DM abundance ($\Omega_{\mbox{\tiny DM}}$) and the abundance of baryons ($\Omega_{\mbox{\tiny B}}$), $\Omega_{\mbox{\tiny B}}/\Omega_{\mbox{\tiny DM}}\sim 1/5$ by relating primordial asymmetries in the two sectors \cite{Komatsu:2010fb}.
 
Like in the case of baryons, if there exists a primordial particle/anti-particle asymmetry in the dark sector, annihilations proceed until depletion of the symmetric part of the population. The final abundance of DM is therefore controlled by the primordial asymmetry, in a contrast to usual freeze-out WIMP scenarios where the relevant quantity is the value of the DM annihilation cross section.  A possible issue in this approach is that 
if the asymmetry in the dark sector is linked to the baryon number, then the observed DM abundance is explained naturally for rather light, {$\mathcal{O}$} (5 GeV) DM mass. 
  
This conclusion changes in the presence of oscillations between DM and anti-DM particles, and in \cite{Cirelli:2011ac} we study this process in detail (see also \cite{Buckley:2011ye,Tulin:2012re}). Oscillations replenish the depleted population of `targets' so that annihilations can re-couple and deplete further the DM/anti-DM abundance. The final DM relic abundance is therefore attained through a more complex history than in the standard aDM case and in a closer similarity to the freeze-out one opening up the parameter space towards heavier DM masses in a natural way.

\section{Theory motivations}
\label{sec:theory}

In \cite{Cirelli:2011ac} we assume that the dark matter particle DM is not its antiparticle $\overline{{\rm DM}}$ and that there exists a primordial asymmetry between the two populations. We then study the evolution of the two populations  in the presence of oscillations generated by a $\Delta ({\rm DM})=2$ mass  term,  $\delta m$. The effect of $\Delta ({\rm DM})=2$ operators is to introduce a mass splitting and mixing between DM and $\overline{{\rm DM}}$, which are no longer mass eigenstates. We will use the generic form of a Hamiltonian
\begin{equation}
{\cal H} =
 \left(
 \begin{array}{cc}
m&\delta m   \\
 \delta m & m    
\end{array}
\right)
\ \ \mbox{where } \ \ 
 \delta m=
  \left\{
 \begin{array}{cc}
\Delta  &  \mbox{ if fermionic DM}\\
\Delta^2/(4M) &     \mbox{ if bosonic DM}
\end{array}
\right.
\end{equation}

The value of the `Majorana' mass $\Delta$ is naturally much smaller than the `Dirac' mass $m$ since $\Delta$ violates a global $U(1)_{\mbox{\tiny DM}}$ symmetry. In our model-independent study, we leave $\delta m$ to be a free parameter which will be scanned over orders of magnitude in the sub-eV range. 
A natural value for this parameter in the fermionic case is obtained from  the dimension-5 operator $ \frac{XXH^{\dagger}H}{\Lambda}$.
After electroweak symmetry breaking  and taking $\Lambda$ at the Planck scale we obtain the see-saw value $\delta m \sim 10^{-6}$ eV.
This value of $\delta m$ turns our to be comparable to the value of the expansion rate  at the time of WIMP freeze out 
and it is therefore expected that it will lead to interesting cosmological effects, as me demonstrate below.

\section{Oscillation + annihilation + scattering formalism}
\label{sec:formalism}

Our aim is to study the evolution in time $t$ of the populations of DM particles ( $n^+$) and their antiparticles $\overline{{\rm DM}}$, ($n^-$),
which are subject to the simultaneous processes of {\em annihilations} ${\rm DM}\, \overline{{\rm DM}} \rightarrow {\rm SM}\, \overline{{\rm SM}}$ (with ${\rm SM}$ being any Standard Model particle), {\em oscillations} ${\rm DM} \leftrightarrow \overline{{\rm DM}}$ and {\em elastic scatterings} ${\rm DM}\, {\rm SM} \rightarrow {\rm DM}\, {\rm SM}$. For definiteness, we assume that particles are initially more abundant than antiparticles, i.e. $n^+ > n^-$.

This problem, in which a coherent process such as oscillations is overlapping with incoherent processes such as annihilations and scatterings is best described by the density matrix formalism, originally developed for the case of neutrino oscillations in the Early Universe~\cite{formalism}. In this approach a $2\times 2 $ matrix is defined with diagonal entries correspond to the individual number densities $n^+$ and $n^-$ and whose off-diagonal entries express the superposition of quantum states ${}^+$ and ${}^-$ originated by the oscillations.  
As is customary, we recast the problem in terms of the co-moving densities $Y^\pm \equiv n^\pm/s$, where $s$ is the total entropy density of the Universe, and we follow the evolution in terms of the dimensionless variable $x= m_{\mbox{\tiny DM}}/T$, where $m_{\mbox{\tiny DM}}$ is the  DM mass and $T$ the temperature. We will therefore work in terms of a {\em comoving number density matrix} 
\begin{equation}
\label{densitymatrix}
\mathcal{Y}(x) = \left( \begin{array}{cc} Y^+(x) & Y^{+-}(x)\\ Y^{-+}(x) & Y^-(x) \end{array} \right)
\end{equation}
(the curly font for $\mathcal{Y}$ will indicate in the following the matrix quantity). We will always be interested in the epoch of radiation domination, during which the Hubble parameter $H(x)=\sqrt{8\pi^3 g_*(x)/90}\, m_{\mbox{\tiny DM}}^2  x^{-2}/M_{\rm Pl}=H_m/x^2$ and $t^{-1} = 2 H(x)$. Here $g_*(x)$ and $g_{*\rm s}(x)$ are the effective relativistic degrees of freedom. 

The evolution equation for the density matrix $\mathcal{Y}$ then reads (with $\  ^\prime = \frac{1 }{x\, H(x)} \times \frac{d\, }{dt} $.)
\begin{eqnarray}
\label{masterequation}
\mathcal{Y}^{\, \prime}(x) & = & -  \frac{i}{x\, H(x)} \Big[\mathcal{H},\mathcal{Y}(x) \Big] \\  \nonumber
&  & - \frac{s(x)}{x\, H(x)} \left( \frac{1}{2} \Big\{ \mathcal{Y}(x), \Gamma_{\rm a}\, \bar{\mathcal{Y}}(x) \, \Gamma_{\rm a}^\dagger \Big\}  -  \Gamma_{\rm a} \, \Gamma_{\rm a}^\dagger \, \mathcal{Y}_{\rm eq}^2 \right) \\  \nonumber
& & - \frac{1}{x\, H(x)} \Big\{ \Gamma_{\rm s}(x), \mathcal{Y}(x) \Big\}. 
\end{eqnarray}

On the right hand side, the first term accounts for oscillations, the second for annihilations and the third for elastic scatterings, with the corresponding operators defined as (for more detail see \cite{Cirelli:2011ac}):
\begin{equation}
\label{HandGammas}
\mathcal{H} =  \left( \begin{array}{cc} m_{\mbox{\tiny DM}} +V(x) + \Delta V(x) & \delta m \\ \delta m & m_{\mbox{\tiny DM}}  +V(x) \end{array} \right),~\Gamma_{\rm s} =  \left( \begin{array}{cc} \gamma_{\rm s} & 0 \\ 0 & \gamma_{\rm s} \end{array} \right)~ {\rm and} ~ \Gamma_{\rm a} \, \Gamma_{\rm a}^\dagger = \langle \sigma v \rangle \, \rm{I}.
\end{equation}
where $\langle \sigma v \rangle$ is the thermally averaged annihilation cross section and $\Delta V$ represents the effective energy shift of DM versus  $\overline{\rm DM}$ induced by the baryon asymmetry of the medium and $\gamma_s$ is the elastic scattering cross section. Both $\Delta V$ and $\gamma_s$ depend on the scattering cross section of DM with ordinary matter and we parametrize them through the ratio with the standard 'weak interaction' values. The direct detection experiments constrain this ratio to be $\leq 10^{-2}$, and we will use two values $\xi=0$ and $\xi=10^{-2}$ in the following. The common terms on the diagonal of $\mathcal{H}$, Dirac mass term $m_{DM}$ and an effective matter potential $V$ do not affect oscillations.

In  eq.~(\ref{masterequation}), $Y_{\rm eq}$ denotes an equilibrium comoving density $Y_{\rm eq} = \frac{45}{2 \pi^4}\left(\frac{\pi}{8}\right)^{1/2}\frac{g}{g_{*\rm s}}x^{3/2} e^{-x}$, where $g$ is the number of internal degrees of freedom and $\bar{\mathcal{Y}}$ is the charge-conjugated matrix of $\mathcal{Y}$, i.e. the same quantity as the latter but with the role of particles and antiparticles flipped: $\bar{\mathcal{Y}} =  {\rm CP}^{-1} \cdot \mathcal{Y}\cdot {\rm CP}$, where ${\rm CP} = i \sigma_2 ={\tiny \left( \begin{array}{cc}0 & 1\\ -1 & 0\end{array} \right)}$. The initial conditions read $Y^\pm_0 \equiv Y^\pm (x_0) = Y_{\rm eq}(x_0) \, e^{\pm \xi_0}$ and $Y^{+-}(x_0) = Y^{-+}(x_0) =0$, at an initial time $x_0$. It is also useful to introduce the  parameter $\eta_0=Y^+_0-Y^-_0$, which represents the initial DM -- $\overline{{\rm DM}}$ asymmetry and is related to  $\xi_0$ as $\xi_0 = {\rm arcsinh}(\eta_0/(2 Y_{\rm eq}(x_0)))$.

The interplay of the coherent and incoherent processes (annihilations and scatterings) can thus be thoroughly followed by using the full density matrix formalism 
\footnote{\em{In a recent work \cite{Tulin:2012re} authors derive the density matrix equations for DM oscillations and freeze-out from first principles using non equilibrium field theory, and arrive to equations which in practical applications often lead to similar results as ours, although in some cases can differ. A paper containing a thorough discussion of these issues is in preparation.}}. We summarize our results by showing the contour lines corresponding to the correct DM abundance in the $(m_{\mbox{\tiny DM}},\sigma_0)$ plane in Fig.~\ref{paramspace}. By changing the values of $\delta m$ (and in general also $\eta_0$, which is not shown here) we can open much more of the parameter space, towards larger $m_{\mbox{\tiny DM}}$ and larger $\sigma_0$. 

As oscillations symmetrize the dark sector allowing for annihilations to resume, the parameter space presented above is subject to the usual constraints on annihilating dark matter. The most stringent constraints are set by the indirect DM searches in the present epoch provided by the {\sc FERMI-LAT} \cite{FERMidwarfs,haloconstraints} and by the observation of the Galactic Center halo with the {\sc H.E.S.S.} telescope~\cite{Abramowski:2011hc}, as well as  from considering the effect on the generation of the CMB anisotropies at the epoch of recombination (at redshift $\sim1100$) and their subsequent evolution down to the epoch of reionization \cite{CMBconstraints}. We show these constraints superimposed on the available parameter space, in Fig.~\ref{paramspace}. 

\begin{figure}[!t]
\begin{center}
\includegraphics[width=0.49 \textwidth]{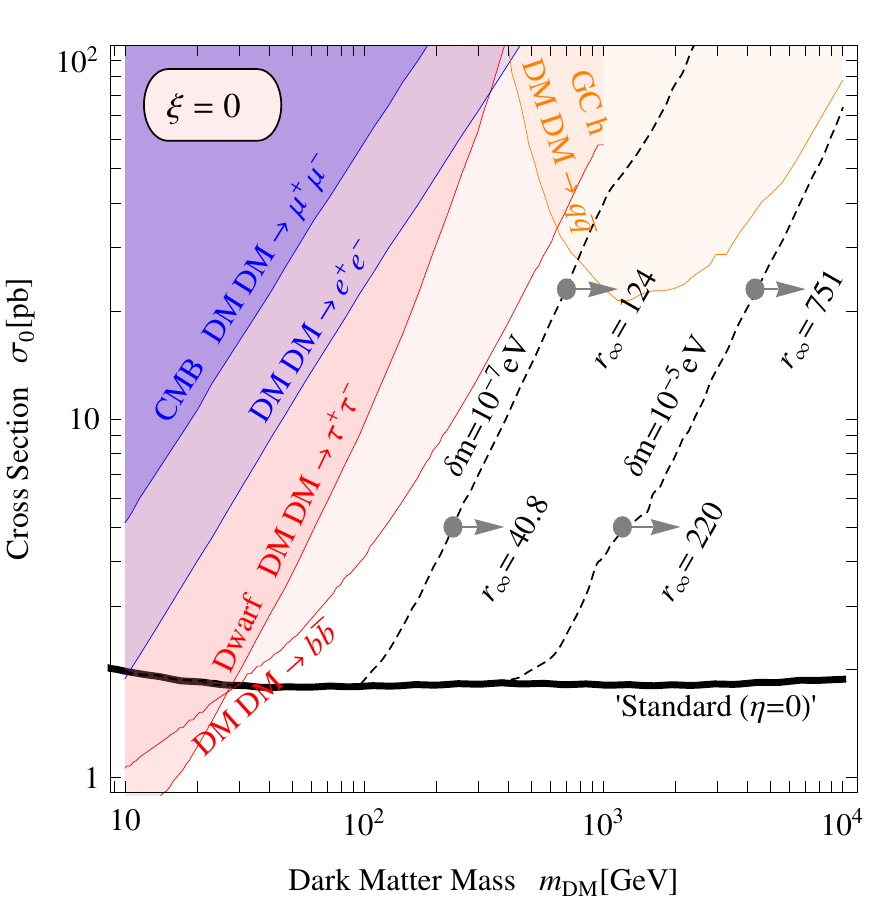}
\includegraphics[width=0.49 \textwidth]{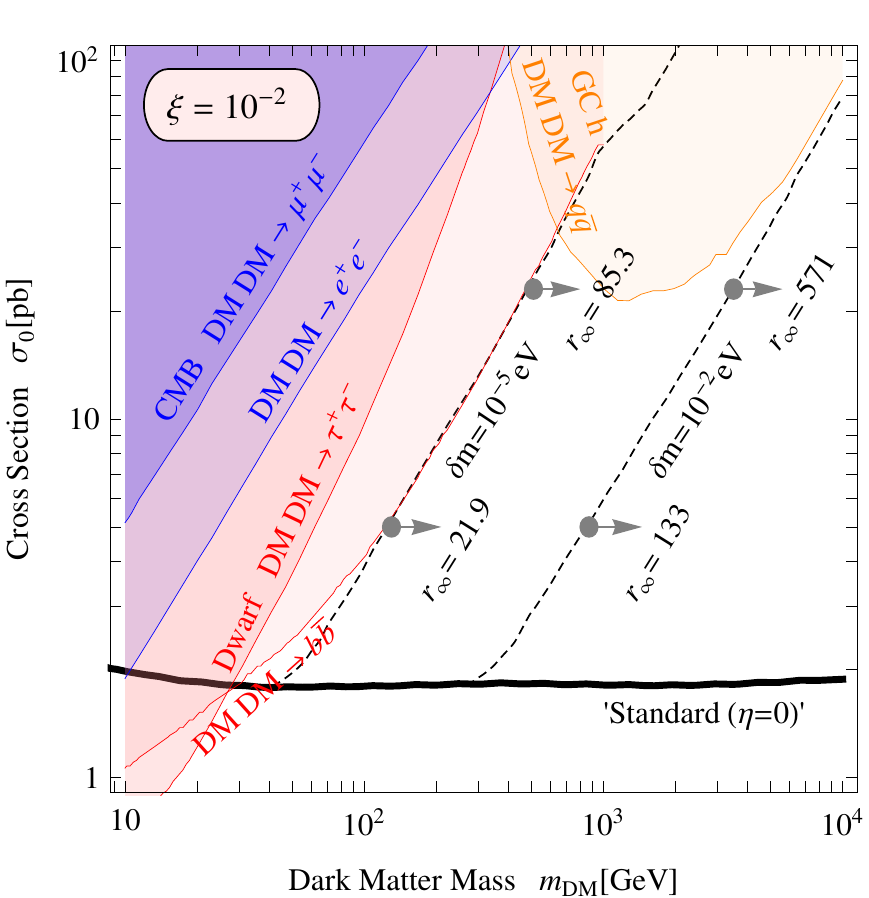}
\caption{\em \small \label{paramspacezoom} Summary plots of the parameter space showing the constraints. The dotted lines mark the contours along which a correct DM abundance 
can be obtained. {\bf \em Left}: Oscillations and annihilations only ($\xi = 0$). {\bf \em Right}: Including elastic scatterings ($\xi = 10^{-2}$). In both panels we assume an initial asymmetry $\eta_0 = \eta_{\mbox{\tiny B}}$ and we show two indicative values of the oscillation parameter $\delta m$. The solid  black line at the bottom represents the standard case ($\eta = 0, \delta m = 0$). The shaded blue regions are excluded by CMB constraints, the shaded pink ones are disfavored by gamma ray observations with {\sc FERMI} and the orange ones by observations with {\sc H.E.S.S.} (see text). The white areas above the solid black line are allowed.}
\label{paramspace}
\end{center}
\end{figure}

\section{Summary}

Oscillations arise naturally in aDM models and we have studied the impact of adding oscillations between DM and $\overline{{\rm DM}}$ particles on the scenario of Asymmetric Dark Matter (see also \cite{Buckley:2011ye,Tulin:2012re}). We found in particular that a typical WIMP with a mass at the EW scale ($\sim$ 1 TeV) having a primordial asymmetry of the same order as the baryon asymmetry, naturally gets the correct relic abundance if the $\delta m$ mass term is in the $\sim$ meV range. This turns out to be a natural value for fermonic DM arising from the higher dimensional operator $H^2\, {\rm DM}^2/\Lambda$ where $H$ is the Higgs field and $\Lambda \sim ~M_{\rm Pl}$. 

We have applied the matrix formalism to explore the phenomenologically available parameter space, by varying the dark matter mass $m_{\mbox{\tiny DM}}$, the annihilation cross section $\sigma_0$, the primordial asymmetry in the DM sector $\eta_0$ and the mass difference $\delta m$, for two discrete choices of the parameter $\xi$ that sets the strength of the elastic scatterings between DM and the plasma.
The quantitative results are displayed in Fig~\ref{paramspacezoom}, together with the most relevant constraints defining the still allowed regions.

\section*{Acknowledgments}
The author thanks the organizers of the 24th Rencontre de Blois for a very interesting meeting as well as M. Cirelli, P. Panci and G. Servant for comments on the manuscript.

%

\end{document}